\documentclass[10pt,twocolumn]{IEEEtran}
\usepackage{amsmath}
\usepackage{amssymb}
\usepackage{amsfonts}
\usepackage{bbm}

\newcommand{\Rmnum}[1]{\expandafter\@slowromancap\romannumeral #1@}
\usepackage{graphicx}
\usepackage{epsfig}
\usepackage{float}
\usepackage{subfigure}
\usepackage[sort]{cite}
\usepackage{xcolor}
\usepackage{enumerate}
\usepackage{extarrows}
\usepackage{algorithmic}
\usepackage[lined,ruled,linesnumbered]{algorithm2e}
\usepackage{tensor}
\usepackage{balance}
\usepackage[draft,bookmarks=false]{hyperref}
\begin{document}

\title{Sparsity-Aware Event-Driven Impulse Radio Transceivers for Reliable Neuromorphic Inference}

\author{\IEEEauthorblockN{Zhengzhong Guan${}^{\ast}$, Jiaying Li${}^{\ast}$, Kanghua Li${}^{\ast}$, Bojun Cheng${}^\ast$, and Hong Xing${}^{\ast\dagger}$}\\
	\IEEEauthorblockA{${}^\ast$ The Hong Kong University of Science and Technology (Guangzhou), Guangzhou, China \\
    ${}^\dagger$ The Hong Kong University of Science and Technology, HK SAR, China \\
		E-mails:~\{zguan807, jli989, kli413\}@connect.hkust-gz.edu.cn, bocheng@hkust-gz.edu.cn,~hongxing@ust.hk
  }}

\maketitle
\thispagestyle{empty}
\IEEEpeerreviewmaketitle

\newtheorem{definition}{\underline{Definition}}[section]
\newtheorem{fact}{Fact}
\newtheorem{assumption}{Assumption}
\newtheorem{theorem}{\underline{Theorem}}[section]
\newtheorem{lemma}{\underline{Lemma}}[section]
\newtheorem{proposition}{\underline{Proposition}}[section]
\newtheorem{corollary}[proposition]{\underline{Corollary}}
\newtheorem{example}{\underline{Example}}[section]
\newtheorem{remark}{\underline{Remark}}[section]

\newcommand{\mv}[1]{\mbox{\boldmath{$ #1 $}}}
\newcommand{\mb}[1]{\mathbb{#1}}
\newcommand{\Myfrac}[2]{\ensuremath{#1\mathord{\left/\right.\kern-\nulldelimiterspace}#2}}
\newcommand\Perms[2]{\tensor[^{#2}]P{_{#1}}}

\begin{abstract}
The growing number of Internet-of-Things (IoT) based artificial intelligence (AI) applications deployed at resource-constrained network edge call for ultra-reliable and low-latency data processing pipelines from distributed front-end sensors to remote inference units. Meanwhile, brain-inspired neuromorphic computing featuring spiking neural networks (SNNs) have arisen as a new paradigm for energy-efficient AI inference. However, significant energy and time expenses incurred in high-complexity transceivers that combat fading and multi-user interference hinder implementations of multi-user neuromorphic inference for edge intelligence. To address this challenge, we consider in this paper a broadband multi-user remote inference system that integrates event-based sensing and time-hopping (TH) on-off keying (OOK) based ultra-wideband (UWB) communications for reliable neuromorphic inference. Specifically, we propose a novel \emph{two-timescale repetition coding} that leverages intra-frame pulse sparsity for low-latency repetition. We also develop two neuromorphic inference schemes based on: (i) digital spike encoding that recovers each pixel of the event-frame by threshold-adaptive detection via an SNN based sparsity estimator; and (ii) analog spike encoding that converts noisy correlator outputs at the receiver into analog-valued inputs for end-to-end (E2E) classification. Finally, numerical results validate the effectiveness of the proposed coding schemes, and reveal a signal-to-noise ratio (SNR)-dependent performance crossover between the two inference schemes, indicating that analog spike encoding based schemes are preferable with mild or high SNR while digital spike encoding based schemes remain robust in low SNR regime.
\end{abstract}

\section{Introduction}
\label{sec:introduction}
The proliferation of Internet-of-Things (IoT) devices and edge intelligence applications envision unprecedented deployment of artificial intelligence (AI) model inference in resource constrained edge devices~\cite{xing2023task,liu2024distributed_snn}. Apart from one streamline of work dedicated to improving edge inference performance via lightweight models by compression \cite{lamaakal2025survey} or distillation \cite{sun2024cooperative}, another streamline of efforts focuses on brain-inspired model inference that relies on neuromorphic computing processors operating on binary spikes \cite{8887558,yang2022lead_federated}. The latter paradigm, together with underlying spiking neural networks (SNNs) \cite{roy2019snn_survey}, enables low-power and real-time inference in latency-critical applications such as autonomous driving \cite{maqueda2018event}, neuromorphic tactile perception \cite{ortone2026tactile} and embodied intelligence \cite{bartolozzi2022embodied}, among which event-based data processing pipelines can further reduce data volume by over $90\%$ \cite{lichtsteiner2008dvs}.
 
In many practical IoT deployments, however, the event-triggered sensor and the inference processor are not co-located, thus entailing transmission of event streams over a wireless link. Unlike ideal communications often assumed in neuromorphic systems~\cite{amir2017low,christensen2023neuromorphic,shrestha2022review}, wireless channels introduce multipath fading and multi-user interference that can distort event-based data transmission and degrade downstream SNN inference, motivating reliable channel-aware transceiver designs that are compatible with event-based data processing pipelines for neuromorphic inference. Ultra-wideband (UWB) time-hopping (TH)  based multiple-access provides a natural multi-user modulation scheme to accommodate such low-power and low-latency transmissions thanks to its impulse radio (IR) waveform and TH spread-spectrum based code-division multiple access (CDMA)~\cite{scholtz1993multiple,win2000uwb,
cassioli2002uwb_gaussian}. Moreover, very recent efforts have started to tailor IR-based wireless architectures to neuromorphic inference for event-driven semantic communications~\cite{chen2023neuromorphic} and neuromorphic split computing~\cite{chen2023nisac,chen2024wakeup}, in which an SNN encoder at the sensor side extracts spike-encoding features that are directly transmitted as IR pulses to a remote SNN decoder for inference.
 
Existing transceiver designs for SNN-based inference mainly follow two design approaches: the first one leverages data reconstruction based methods that demodulate and decode the received signal to recover original (binary) spike representations prior to SNN-based inference in the sequel~\cite{wu2025multilevel}; and the second one is task-oriented, by which remote inference units avoid hard decisions on bits, but feed the received signal directly into SNN for inference, pursuing end-to-end (E2E) performance metrics, e.g., test accuracy for classification tasks, rather than bit-level fidelity~\cite{wang2025snnsc,chen2023neuromorphic,chen2023nisac,chen2024wakeup}. 
 
However, there is a lack of principled physical-layer transceiver designs and systematic guidelines for choosing between these two design streams. On one hand, most existing solutions do not explicitly translate event-driven sparsity of spikes into IR based modulation and/or channel coding judiciously designed to accommodate latency requirements and highly imbalanced transmission-bit priors between zero-bit and one-bit \cite{liu2024spectrum, liu2024symbol}. On the other hand, existing works either simplify multiple access by adopting orthogonal transmissions, such as orthogonal frequency-division multiple access (OFDMA)~\cite{wu2025multilevel}, or consider simplified channel models that do not capture the rich multipath propagation inherent in practical UWB environments~\cite{chen2023neuromorphic,chen2023nisac, chen2024wakeup}. To fill in these gaps, in this paper, we propose a sparsity-aware IR transceiver design leveraging TH on-off keying (OOK) modulation and two-timescale repetition coding for reliable neuromorphic inference.

Our contributions are summarized as follows. 
1) For the IR transmitter, to exploit the inherent sparsity of neuromorphic sensors' event-based data, we adopt TH-OOK modulation for each user to significantly save power while reducing chances of multi-user's pulse collision. 
2) We propose a novel sparsity-aware two-timescale repetition coding scheme that exploits low pulse density inherent in OOK-modulated sparse events to enable \emph{intra-frame} repetitions within one transmission frame without excessive delay,  thereby improving coding efficiency.
3) We develop two neuromorphic inference frameworks, namely digital spike encoding based scheme, which fundamentally separates model-based event-frame reconstruction from SNN based object recognition, and an analog spike encoding based scheme that directly performs object recognition from analog input in an E2E fashion.
4) Our numerical results reveal insights that provide useful guidance for effective selection of the above two schemes based on signal-to-noise ratio (SNR) conditions of wireless channels.

% ============================================================
% Section II: System Model and Problem Formulation
% ============================================================

\section{System Model}
\label{sec:system_model}

\begin{figure}[!t]
\centering
\includegraphics[width=\columnwidth]{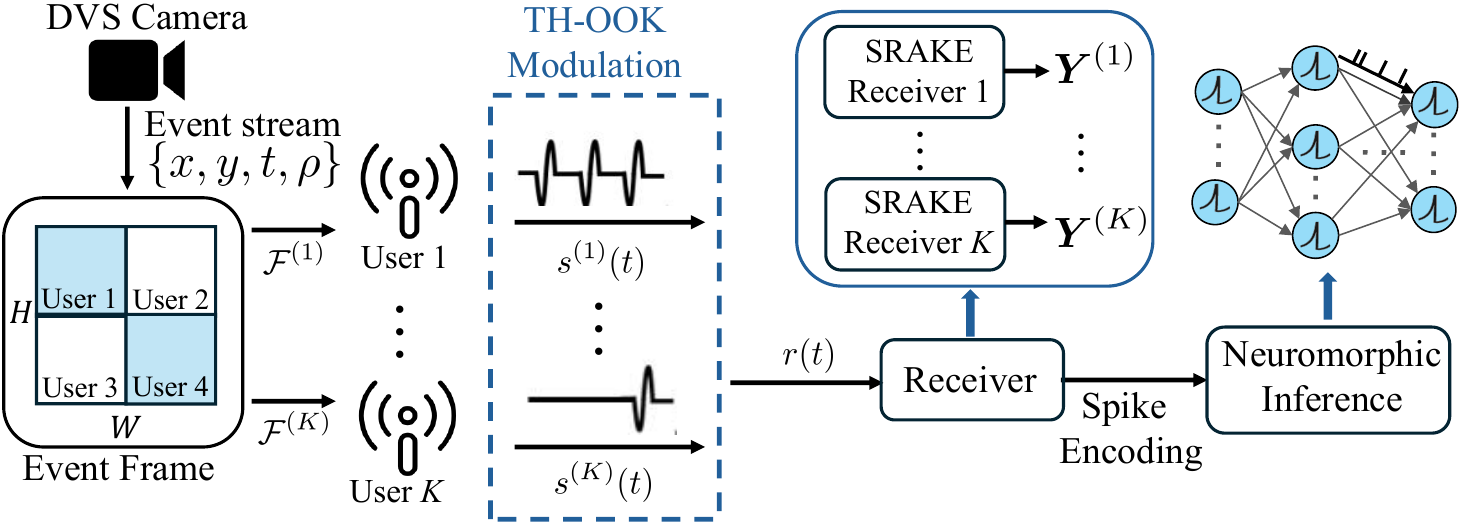}
\caption{Illustration of the system model.}
\vspace{-0.25in}
\label{fig:system_overview}
\end{figure}

We consider a wireless event-driven communication system for real-time object recognition, as shown in Fig.~\ref{fig:system_overview}. 
The system comprises $K$ spatially distributed users, each equipped with a single-antenna IR transmitter (Tx), sending data within its field of view captured by a dynamic vision sensor (DVS) camera, and a remote single-antenna IR receiver (Rx) that performs real-time object recognition using an integrated neuromorphic computing unit.

% ============================================================
% Section II.A: Sensing Model
% ============================================================
\subsection{Sensing Model}
\label{subsec:sensing}
At the sensing side, the users' DVS cameras asynchronously capture brightness changes as streams of sparse events~\cite{lichtsteiner2008dvs}. 
Each event is characterized by a 4-tuple $(x, y, t, \rho)$, where $(x,y)$ denotes the spatial coordinate for a pixel of interest; $t$ is the timestamp; and $\rho \in \{-1, 1\}$ indicates polarity (brightness decreases or increases). 
To preserve polarity information, each pixel in $(x,y)$ is encoded using 2 bits: ``10'' for decrease ($\rho=-1$), ``01'' for increase ($\rho=1$) and ``00'' for no change in brightness. 
As a result, events are accumulated over a time duration $\Delta T$ to form an event frame $\mathcal{F} \in \{0,1\}^{H \times W \times 2}$ with spatial resolution $H \times W$.

Assuming that the visual data captured by users in $\{ 1, \ldots, K\}\triangleq [K]$ are non-overlapping, we have a complete event frame composed of $K$ disjoint sets of tiles $ \in\{0,1\}^{H\times W\times 2}$, with each set corresponding to user $k\in[K]$, where $\bar{H} = H/\sqrt{K}$ and $\bar{W} = W/\sqrt{K}$. 
Each set $\mathcal{F}^{(k)}$ can be vectorized into a bit stream $\{b_n^{(k)}\}_{n=1}^{N_b}$ of length $N_b = 2\bar{H}\bar{W}$, where each bit $b_n^{(k)} \in \{0,1\}$ denotes the $n$-th position in the bit stream.

% ============================================================
% Section II.B: Transmission Model
% ============================================================

\subsection{Communication Model}
\label{subsec:transmission}

\subsubsection{IR Transmitter}
\label{subsubsec:tx_model}
We adopt TH-OOK modulation. Specifically, each bit in the bit stream of each user is transmitted in one shared transmission frame of duration $T_f$ that is divided into $N_h$ time-hopping chips each of duration $T_c$, such that $T_f=N_hT_c$. To accommodate multiple access, each user is assigned a unique pseudo-random sequence $\{c_m^{(k)}\}$, $k\in[K]$, with $c_m^{(k)}\in\{0,1,\ldots, N_h-1\}$ denoting the chip index selected by user $k$ in the $m$-th transmission frame. In addition, the ultra-wideband pulse transmitted by each user in each transmission frame is modulated using OOK. That is, when $b_n^{(k)}=1$, user $k\in[K]$ transmits a monocycle pulse in the $c_n^{(k)}$-th chip, and remains silent otherwise.

To enhance reliability against fading and multi-user interference, we employ \emph{inter-frame} repetition coding, where each bit in the bit stream $\{b_n^{(k)}\}_{n=1}^{N_b}$ of user $k\in[K]$ is repeatedly transmitted for $N_f$ consecutive frames with each frame once. The transmitted signal for the $n$-th bit of user $k\in[K]$ at timestamp $t$ is thus given by \cite{win2000uwb}:
\begin{equation}
s_n^{(k)}(t) = \sqrt{E_s} \sum_{j=0}^{N_f-1} b_n^{(k)} \, 
w\!\left(t - jT_f - c_{(n-1)N_f+j}^{(k)}T_c\right),
\label{eq:th_ook_signal}
\end{equation}
where $E_s$ denotes the energy per symbol; and $w(t)$ represents the transmitted monocycle waveform such that $\int_{-\infty}^{\infty} w^2(t)\,dt = 1$.
\subsubsection{Channel Model}
\label{subsubsec:channel_model}
We assume a quasi-static multi-path block fading for UWB channels, where the channel remains constant over a block of (e.g., tens of) transmission frames for each user, and varies independently across blocks and users~\cite{cassioli2002uwb_gaussian}.
Specifically, the multi-path channel impulse response for user $k\in[K]$ is expressed as:
\begin{equation}
h^{(k)}(t) = \sum_{l=1}^{L_c} \sum_{m=1}^{M_l} \alpha_{l,m}^{(k)} \,
\delta\!\left(t - T_l^{(k)} - \tau_{l,m}^{(k)}\right),
\label{eq:multi-path_channel}
\end{equation}
where $T_l^{(k)}$ and $\tau_{l,m}^{(k)}$  denote the delay of cluster $l$ and the $m$-th ray in cluster $l$, \(m\in[M_l]\), \(l\in[L_c]\), respectively.
The corresponding complex gain is $\alpha_{l,m}^{(k)}$, which follows Nakagami-$m$ distribution in magnitude and uniform distribution in phase.

\subsubsection{IR Receiver}
\label{subsubsec:rx_model}
The received signal of the common transmission frame corresponding to the $n$-th bit of all users is:
\begin{equation}
r_n(t) = \sum_{k=1}^{K} \int_{-\infty}^{\infty} s_n^{(k)}(\tau) h^{(k)}(t-\tau) \, d\tau + n(t),
\label{eq:received_signal_per_bit}
\end{equation}
where $n(t)$ is the additive white Gaussian noise (AWGN), denoted by $n(t)\sim \mathcal{N}(0,N_0/2)$. 
Plugging~\eqref{eq:th_ook_signal} and~\eqref{eq:multi-path_channel} into~\eqref{eq:received_signal_per_bit}, we obtain
\begin{equation}
\begin{aligned}
r_n(t) = &\sum_{k=1}^{K} \sqrt{E_s} \sum_{j=0}^{N_f-1} b_n^{(k)} \sum_{l=1}^{L_c} \sum_{m=1}^{M_l} \alpha_{l,m}^{(k)} \\
&\times w\!\left(t - jT_f - c_{(n-1)N_f+j}^{(k)}T_c - \tau_{\ell}^{(k)}\right) + n(t),
\end{aligned}
\label{eq:received_signal_expanded}
\end{equation}
where $\tau_\ell^{(k)} \triangleq T_l^{(k)} + \tau_{l,m}^{(k)}$. 
We assume that UWB channel parameters (such as $\tau_{l,m}^{(k)}$'s and $\alpha_{l,m}^{(k)}$'s) can be obtained by effective estimation~\cite{ieee802154_2020,lottici2002channel}, and thus accessible by the receiver.

To combat multi-path fading, the Rx employs selective RAKE (SRAKE) combining~\cite{cassioli2007rake}, which selects the $L$ strongest paths by power $|\alpha_{l,m}^{(k)}|^2$.
For the $\ell$-th selected path, we denote its complex gain by $\alpha_{\ell}^{(k)}$, and the SRAKE Rx correlates the received signal with $LK$ monocycle waveform templates with each corresponding to one of the $L$ strongest paths of one user in each repeated frame.  
Thus, the correlator output for the $n$-th bit of user $k\in[K]$ corresponding to path $\ell\in[L]$ for frame $j \in \{0,1,\ldots, N_f-1 \} \triangleq[N_{f-}]$ is given by
%{-0.1cm}
\begin{equation}
\begin{aligned}
Y_{n,j,\ell}^{(k)} &= \int_{jT_f}^{(j+1)T_f} r_n(t)\\
&\times w\!\left(t - jT_f - c_{(n-1)N_f+j}^{(k)}T_c - \tau_\ell^{(k)}\right) dt \\
&= \sqrt{E_s} \, b_n^{(k)} \alpha_{\ell}^{(k)} + \xi_{n,j,\ell}^{(k)},
\end{aligned}
\label{eq:rake_correlator}
\end{equation}
where $\xi_{n,j,\ell}^{(k)}$ represents the aggregate interference-plus-noise. Since the interference plus noise aggregates $L_{\text{total}}=\sum_{l} M_{l}$ resolvable components of all users\footnote{for example, $M_{l}$ = 20 and $l$ = 5  under IEEE 802.15.4a channel model (CM1) }, $\xi_{n,j,\ell}^{(k)}$ can be approximated by Gaussian according to the central limit theorem, i.e., $\xi_{n,j,\ell}^{(k)} \sim \mathcal{N}(0,(\sigma_{n,\ell}^{(k)})^2)$~\cite{win2000uwb}.

Finally, the SRAKE Rx adopts maximal ratio combining (MRC) for all $L$ paths, yielding a frame-level statistic as

\begin{equation}
Y_{n,j}^{(k)} = \sum_{\ell=1}^{L} \beta_{\ell}^{(k)} Y_{n,j,\ell}^{(k)},
\label{eq:mrc_combining}
\end{equation}
where $\beta_{\ell}^{(k)} = (\alpha_{\ell}^{(k)})^* / (\sigma_{n,\ell}^{(k)})^2$  is the combining weight corresponding to path $\ell$, \(n\in[N_b]\), \(k\in[K]\) and \(j\in[N_{f-}]\).

\subsection{Learning Model}
\label{subsec:learning}
An SNN-based neuromorphic computing unit is embedded into the Rx to perform object recognition over $C$ classes. The joint statistic of all $N_f$ received frames corresponding to the $n$-th bit of user $k\in[K]$, \(n\in[N_b]\), $j\in[N_{f-}]$ is then rearranged into a tensor $\mv Y\in\mb{R}^{H\times W\times 2\times N_f}$ with each $\mv Y_j\in\mb{R}^{H \times W\times 2}$, reshaped in the reverse way how the event frame $\mathcal{F}$ is partitioned and vectorized. The input spikes of the SNN are then generated by an encoding module, denoted by \(\mathcal{E}(\mv Y):\mb{R}^{H\times W\times 2 \times N_f}\mapsto\{0,1\}^{H\times W\times 2}\). In this work, we consider two neuromorphic encoding modes at the receiver--SNN interface: 1) digital detection-based encoding; 2) analog detection-based encoding, which will be detailed in Section~\ref{sec:training}.

The SNN consists of $D$ layers of leaky integrate-and-fire (LIF) neurons~\cite{roy2019snn_survey,yang2022lead_federated} and processes inputs over $T$ consecutive time steps, with one binary output spike at each time step, for rate-decoding based classification.

% ============================================================
% Section III: Proposed Communication Architecture
% ============================================================
\section{Sparsity-Aware Two-Timescale Repetition Coding for Multi-User IR Transmission}
\label{sec:architecture}

% Figure: inter-frame vs intra-frame Repetition
\begin{figure}[!t]
\centering
\includegraphics[width=\columnwidth]{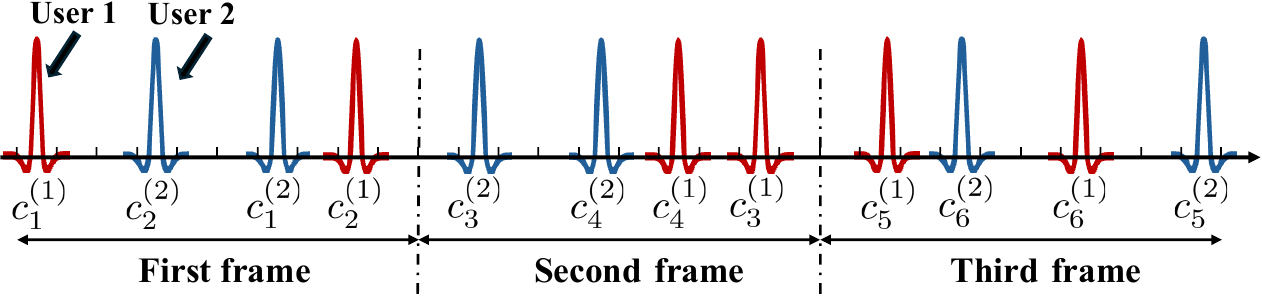}
%{-0.2cm}
\caption{{Illustration of the proposed two-timescale repetition coding for two users' IR transmissions with $N_f=3$ and $N_p=2$.}}
\vspace{-0.2in}
%{-0.2cm}
%{-0.5cm}
\label{fig:repetition_schemes}

\end{figure}

% ============================================================
% Section III.A: intra-frame Repetition Coding
% ============================================================
Since the event-frame data is highly sparse due to the event-driven nature, the activation rate $p = \Pr(b_n^{(k)}=1)$ for the bit stream of each user in each event frame is far less than 0.5. As a result, TH-OOK modulation leaves the majority of chips unoccupied, compared to  PPM, where each bit encoded by a pulse occupies a chip regardless of its value. Such sparsity-induced low pulse density creates a large number of idle chips within each frame, significantly reducing inter-user interference due to pulse collisions. Exploiting this observation, we propose a novel sparsity-aware two-timescale repetition coding that introduces \emph{intra-frame} on top of inter-frame repetition coding, thus achieving $N_s = N_f N_p$ total repetitions using only $N_f$ frames. Specifically, bit $n\in[N_b]$ of user $k\in[K]$, $b_n^{(k)}$, is transmitted $N_f$ times in the frame level and $N_p$ times in the pulse level within each frame. An example is shown in Fig.~\ref{fig:repetition_schemes}.

The transmitted signal for $b_n^{(k)}$, $n\in[N_b]$, $k\in[K]$, is recast as
%{-0.1cm}
\begin{equation}
\begin{aligned}
s_{n}^{(k)}(t) = & \sqrt{E_s} \sum_{j=0}^{N_f-1} \sum_{i=0}^{N_p-1} b_n^{(k)}
\\&\times w(t - jT_f - c_{(n-1)N_s+jN_p+i}^{(k)}T_c),
\end{aligned}
\label{eq:intra-frame_signal}
\end{equation}
 where the pseudo-random index of user $k$ changes by both timescale repetition. At the Rx, the SRAKE first correlates the received signal $r_n(t)$ with a modified waveform template $\sum_{i=0}^{N_p-1}w(t - jT_f - c_{(n-1)N_s+jN_p+i}^{(k)}T_c - \tau_{\ell}^{(k)})$ for the $\ell$-th selected path as in \eqref{eq:rake_correlator}, $\ell\in[L]$, $k \in [K]$, and $j \in [N_{f-}]$, ending up with the correlator output given by
\begin{equation}
Y_{n,j,\ell}^{(k)} = \sqrt{E_s} \, N_p b_n^{(k)} \alpha_{\ell}^{(k)} + \xi_{n,j,\ell}^{(k)},
\label{eq:intra-frame_rake}
\end{equation}
where $\xi_{n,j,\ell}^{(k)}$ represents the aggregate interference-plus-noise.

Next, the SRAKE Rx adopts MRC accordingly to obtain a frame-level statistic \(Y_{n,j}^{(k)}\), \(n\in[N_b]\), \(k\in[K]\), \(j\in[N_{f-}]\), before passing it into the subsequent spike encoding module for SNN-based inference that will be elaborated on in Section~\ref{sec:training}.

% ============================================================
% Section IV: Training Strategies for Neuromorphic Receiver
% ============================================================
\section{SNN-based Neuromorphic Inference Schemes}
\label{sec:training}

In this section, we elaborate on two SNN-based inference schemes that the neuromorphic computing unit in the Rx adopts, based upon two different encoding modes introduced in Section~\ref{subsec:learning}.
% ============================================================
% Section IV.A: Separated Training with Adaptive Detection
% ============================================================

\subsection{Digital Spike Encoding based Inference}
\label{subsec:separated}
This inference scheme mainly encompasses three stages: the first stage focuses on sparsity estimation; the second stage detects each user's bit stream using majority voting via adaptive thresholding; and the third stage makes inference from the reconstructed digital event frame for classification. 
% Figure: Training Schemes Comparison
\begin{figure}[t]
\centering
\includegraphics[width=\columnwidth]{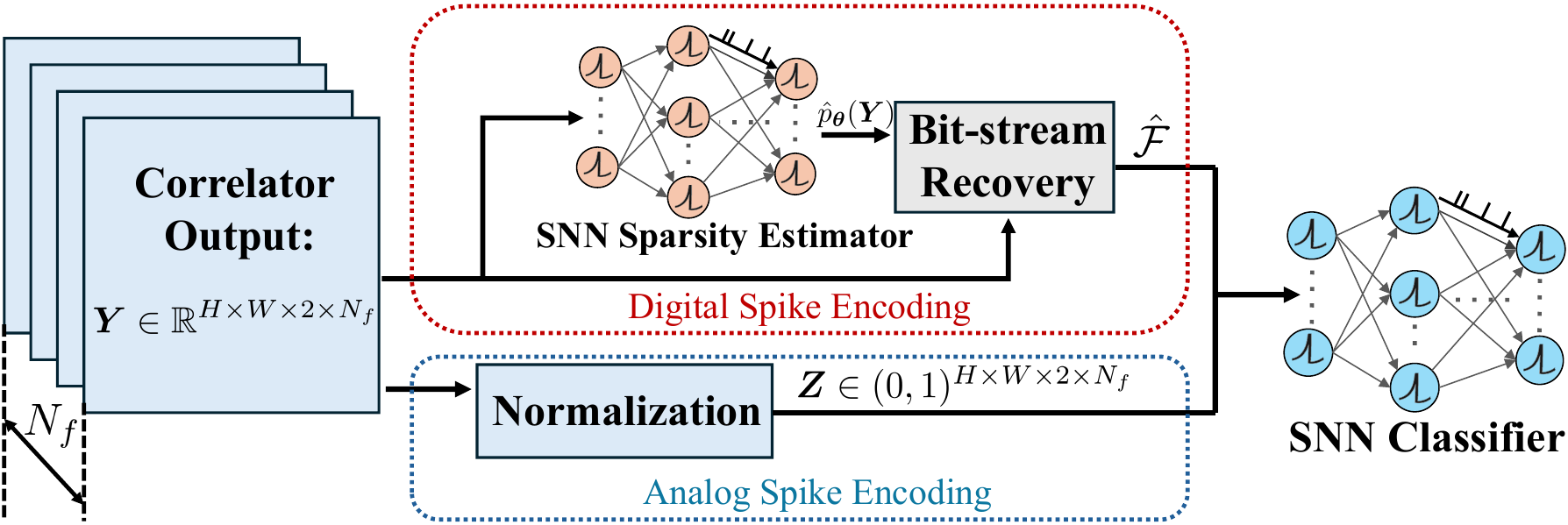}
\caption{Illustration of the digital spike encoding based and analog spike encoding based inference schemes.}
\vspace{-0.2in}
\label{fig:training_schemes}
\end{figure}

For bit-stream recovery in a single frame,
\begin{align}
\tilde{b}_{n,j}^{(k)} = 
\begin{cases}
1, & \text{if } Y_{n,j}^{(k)} > \eta, \\
0, & \text{otherwise},
\end{cases}
\label{eq:single_frame_detection}
\end{align}
where the decision threshold $\eta$ is derived using the maximum a posteriori (MAP) detection rule, as shown in Proposition~\ref{prop:adjusted_threshold}. Note that we omit in the rest of this section the super/subscripts of frame-level statistics and detected bits for ease of exposition without loss of generality.

\begin{proposition}
\label{prop:adjusted_threshold}
Given the transmission prior \(P(b=1)=p\) and a hyper-parameter $\lambda \geq 1$ that controls decision preferences, the MAP decision threshold in terms of $p$ and $\lambda$  is given by
\begin{align}
\eta(\lambda, p) = \frac{\sigma^2}{\mu} \ln\left(\frac{1-p}{\lambda p}\right) + \frac{\mu}{2}.
\label{eq:adapted_threshold}
\end{align}
\end{proposition}
\begin{IEEEproof}
Since $P(b=1)=p \ll 0.5$ due to event sparsity, in contrast to the equal-probability assumption in conventional OOK, we thus introduce the bias parameter \(\lambda\ge 1\) to favor the detection of ``1'' bits in the MAP rule as follows:
\begin{align}
\frac{P(b=1|Y)}{P(b=0|Y)} = \frac{P(Y|b=1)}{P(Y|b=0)} \cdot \frac{p}{1-p} \mathop{\gtrless}_{\tilde{b}=0}^{\tilde{b}=1}  \frac{1}{\lambda}.
\label{eq:map_rule}
\end{align}

Considering the frame-level statistic $Y$ under the Gaussian approximation~\cite{win2000uwb}, we have
$Y$ $\sim$ $\mathcal{N}(0, \sigma^2)$ if $b = 0$; and $Y$ $\sim$ $\mathcal{N}(\mu, \sigma^2)$ otherwise, where
$\mu$ $=$ $\sqrt{E_s} \, N_p \sum_{\ell=1}^{L} \Myfrac{|\alpha_{\ell}^{(k)}|^2}{(\sigma_{n,\ell}^{(k)})^2}$
and
$\sigma^2$ $=$ $N_p \sum_{\ell=1}^{L} \Myfrac{|\alpha_{\ell}^{(k)}|^2}{(\sigma_{n,\ell}^{(k)})^2}$
denote the mean and the variance of the Gaussian distribution, respectively. Then, substituting the conditional probability density function of $Y$ in the decision rule \eqref{eq:map_rule}, we obtain in \eqref{eq:adapted_threshold} the decision threshold $\eta(\lambda,p)$.
\end{IEEEproof}

Next, based on the MAP decision rule in~\eqref{eq:single_frame_detection}, the Rx recovers the bit $b_{n}^{(k)}$, $n \in [N_b]$, $k\in [K]$, by combining $ \tilde{b}_{n,j}^{(k)}$ in all $N_f$ (assumed to be an odd number) frames via majority voting as below:
\begin{align}
\tilde{b}_n^{(k)} = 
\begin{cases}
1, & \text{if } \sum_{j=0}^{N_f-1} \tilde{b}_{n,j}^{(k)} > \frac{N_f}{2}, \\
0, & \text{otherwise}.
\end{cases}
\label{eq:majority_vote}  
\end{align}
Consequently, the bit error rate (BER) in terms of $\lambda$ and $p$  for bit stream recovery of all users is given by  
\begin{equation}
\text{BER}(\lambda, p) = (1-p) \cdot P_{e | 0}(\lambda, p) + p \cdot P_{e | 1}(\lambda, p),
\label{eq:ber_majority_vote}
\end{equation}
where $P_{e|0}(\lambda, p)$ and $P_{e|1}(\lambda, p)$ denote the false alarm and miss detection probabilities over $N_f$ frames, respectively, given by 
\begin{equation}
\begin{aligned}
P_{e|0}(\lambda, p) &= \sum_{i=m+1}^{N_f} \binom{N_f}{i} P_{\text{FA}}^i (1-P_{\text{FA}})^{N_f-i}, \\
P_{e|1}(\lambda, p) &= \sum_{i=0}^{m} \binom{N_f}{i} (1-P_{\text{MD}})^i P_{\text{MD}}^{N_f-i},
\end{aligned}
\label{eq:ber_b0_b1}
\end{equation}
with $m = (N_f-1)/2$, and
\begin{align}
P_{\text{FA}} &= Q\!\left(\frac{\eta(\lambda, p)}{\sigma}\right), \quad
P_{\text{MD}} = Q\!\left(\frac{\mu-\eta(\lambda, p)}{\sigma}\right),\notag
\end{align}
where $Q(\cdot)$ denotes the $Q$-function.

To minimize BER for digital spike encoding, we solve the following optimization problem:
\begin{align}
\mathop{\mathtt{Minimize}}_{\lambda\ge 1, p\in[0,1]}~~\text{BER}(\lambda, p).
\end{align}
Since the event sparsity varies across different  event frames and is unknown to the Rx, we employ an SNN-based deep neural network parameterized by $\boldsymbol{\theta}$ that takes the received tensor $\mv{Y} \in \mathbb{R}^{H \times W \times 2\times N_f}$ corresponding to  each event frame as input and outputs a sparsity estimate $\hat{p}_{\boldsymbol{\theta}}(\mv{Y}) \in [0,1]$. This neural network is  trained offline against ground-truth sparsity $p$ to minimize the mean-square error (MSE) by supervised learning. 

Finally, substituting $\hat{p}_{\boldsymbol{\theta}}(\mv{Y})$ for $p$ in \eqref{eq:ber_majority_vote}, we solve for $\lambda$ by, e.g., the bisection method, to obtain $\lambda^\star$.
With \(\hat{p}_{\boldsymbol{\theta}}(\mv{Y})\) and \(\lambda^\star\), the bit streams of all users are recovered by \eqref{eq:majority_vote}, and then assembled into an estimated event frame $\hat{\mathcal{F}} \in \{0,1\}^{H \times W \times 2}$, which is fed into the pre-trained SNN at the Rx to perform object recognition.

%{-0.3cm}
% ============================================================
% Section IV.B: End-to-End Training
% ============================================================
\subsection{Analog Spike Encoding based Inference}
\label{subsec:end2end}

Different from the digital spike encoding based inference that relies on high-quality bit-stream recovery, thus a hybrid of model-based and model-free inference, the analog spike encoding proposed herein, as shown in Fig.~\ref{fig:training_schemes}, solicits a fully E2E model-free solution by pre-processing the correlator output tensor into a normalized frame of analog pixel values in the same size.
 
Specifically, we use the sigmoid function to normalize the statistic $Y_{n,j}^{(k)}$, resulting in a soft-decision value $z_{n,j}^{(k)} \in (0,1)$, $k\in[K]$, $n\in[N_b]$, $j\in[N_{f-}]$. The soft-decision values from all $K$ users and $N_b$ bits are rearranged into a tensor $\mv{Z} \in (0,1)^{H \times W \times 2 \times N_f}$. The membrane potential of the $d$-th layer at time step $m \in [N_f T]$ is updated by LIF model:
\begin{align}
\begin{aligned}
 \mv{U}^{(d)}[m] = & \beta \mv{U}^{(d)}[m-1] + \mv{W}^{(d)}\mv{X}^{(d)}[m]\\
                                            & \kern-1pt - \mv{S}^{(d)}[m-1] \zeta,
\end{aligned}\label{eq:lif_general}
\end{align}
where $\mv{U}^{(d)}[m]$ and $\mv{S}^{(d)}[m] = \mathbbm{1}(\mv{U}^{(d)}[m] \geq \zeta)$ denote the membrane potential and binary spike output of the $d$-th layer in the SNN, respectively; $\beta \in (0,1)$ is the decay factor; $\mv{W}^{(d)}$ is the weight matrix of the $d$-th layer; $\zeta$ is the firing threshold; and $\mv{X}^{(d)}[m]$ is the synaptic input at the $m$-th time step, which is given by \cite{rathi2023diet}
\begin{align}
\mv{X}^{(d)}[m] = 
\begin{cases}
\mv{Z}_j, & d=1, \\
\mv{S}^{(d-1)}[m], & d = 2, \ldots, D,
\end{cases}
\label{eq:synapse_input}
\end{align}
where $\mv{Z}_j \in (0,1)^{H \times W \times 2}$, $j \in [N_{f-}]$, is the corresponding slice of $\mv{Z}$ along the last dimension.

% ============================================================
% Section V: Experiments
% ============================================================
\section{Experiments}
\label{sec:experiments}
We consider the gesture classification task with the DVS128 Gesture dataset, which consists of $C = 11$ hand gesture classes, divided into $1176$ and $166$ samples for training and testing, respectively~\cite{amir2017low}. Each gesture sequence is temporally reframed into sequences of event frame with $T = 16$ time steps and spatial resolution of $H \times W = 128 \times 128$ pixels. The event frame $\mathcal{F}$ is composed of $K = 16$ non-overlapping tiles $\mathcal{F}^{(k)}$, $k\in[K]$.
To simulate the UWB multi-path channel specified in \eqref{eq:multi-path_channel}, we adopt IEEE 802.15.4a CM1~\cite{ieee802154_2020} with the second-order Gaussian derivative as the monocycle waveform $w(t)$~\cite{scholtz1993multiple}, and set $T_f=100$\;ns with $T_c=2$\;ns. The SNN-based sparsity estimator employs PLIF neurons~\cite{fang2021plif}, where the $N_f$ frames are treated as SNN time steps and processed through four ``Conv--BN--PLIF--AvgPool'' blocks, and the final sparsity estimate is obtained by applying global average pooling and a sigmoid function to the time-averaged membrane potential. The SNN classifier adopts the DVSGestureNet architecture~\cite{fang2021plif} which is pre-trained on clean data and fine-tuned on channel-corrupted data using arctangent surrogate gradients.

For comparison to the proposed two-timescale digital spike encoding based (referred to as \emph{Intra-OOK-D}, and as \emph{OOK-D} when $N_p=1$) and analog spike encoding based (referred to as \emph{Intra-OOK-A}, and as \emph{OOK-A} when $N_p=1$
) inference, we consider the following two benchmarks: 1) the \textit{PPM-D (or \emph{PPM-A})}~\cite{win2000uwb} based scheme, which encodes both $0$-bit and $1$-bit via pulse position modulation (PPM) using only the inter-frame repetition coding; and 2) the \textit{Baseline}, which directly performs classification on a pre-trained SNN in perfect channel conditions without wireless channel impairments. We evaluate test accuracy by averaging over $500$ Monte-Carlo trials.

\begin{figure}[h]
\centering
\includegraphics[width=2.9in]{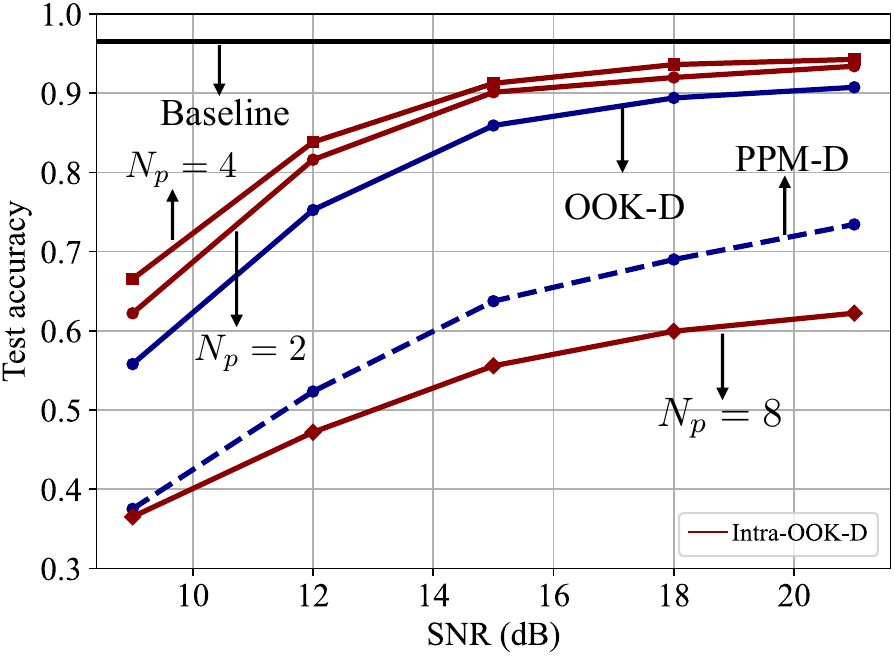}
\caption{Test accuracy versus SNR with $N_f = 9$.}
\label{fig:inter_intra}
\end{figure}

Fig.~\ref{fig:inter_intra} compares test accuracy achieved by different schemes versus received SNR, demonstrating that the intra-frame repetition coding outperforms the inter-frame repetition only coding when $N_p=2$ and $N_p=4$, and that the scheme with $N_p=4$ approaches the baseline under ideal communications in high SNR regime, since larger $N_p$ causes more pulse collisions within one frame.

\begin{figure}[h]
\centering
\includegraphics[width=2.9in]{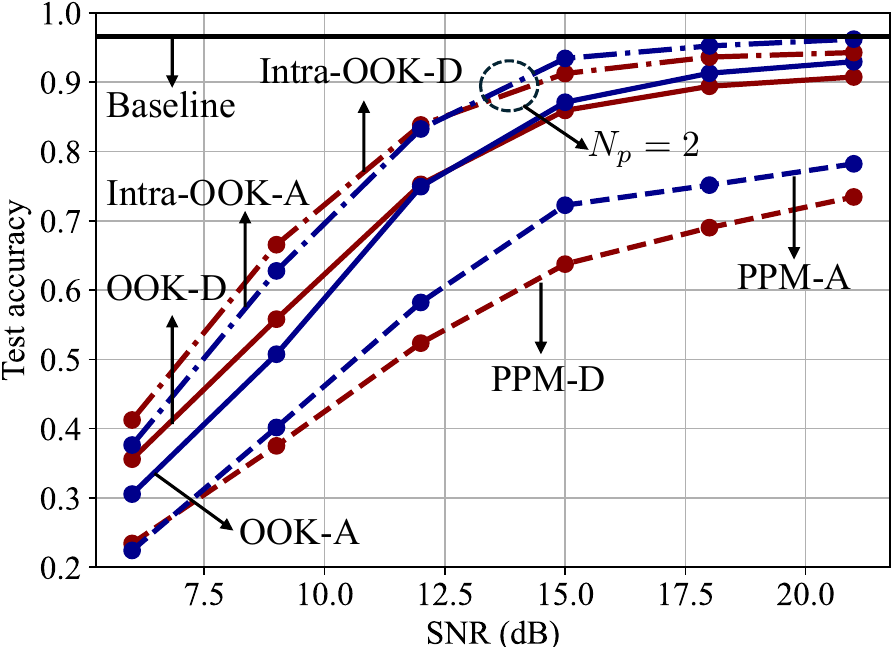}
\caption{Test accuracy versus SNR with $N_f = 9$.}
\label{fig:sp_vs_e2e}
\end{figure}

Fig.~\ref{fig:sp_vs_e2e} illustrates test accuracy versus received SNR by the proposed SNN based inference schemes,
benchmarking the TH-PPM based schemes and the baseline. The sparsity-aware TH-OOK transmissions show striking performance gains over TH-PPM due to significantly reduced pulse collision. Moreover, OOK-D effectively suppresses noise in low SNR regimes, whereas OOK-A excels in learning channel-adapted features in high SNR regimes. This result provides useful guidance for selecting proper inference schemes based on wireless channel conditions.

\section{Conclusion}
\label{sec:conclusion}
In this paper, we proposed a sparsity-aware TH-OOK based UWB transceiver design for event-driven neuromorphic computing over wireless channels. We introduced a two-timescale intra-frame repetition scheme that splits the repetition budgets into frame-level and pulse-level repetitions to improve spectral efficiency, and developed two neuromorphic inference schemes based on digital and analog spike encoding, respectively. Numerical results showed that the proposed repetition coding schemes significantly improved inference accuracy compared to pulse position modulation based benchmarks, and that the two proposed inference schemes exhibited an SNR-dependent performance crossover, providing guidelines for SNR-adaptive remote inference in practical deployment.

\balance
\bibliographystyle{IEEEtran}

\bibliography{main}

\end{document}